# The why, what, and how of AI-based coding in scientific research

Tonghe Zhuang, Zhicheng Lin*

Tonghe Zhuang
Tonghe.Zhuang@psychiat.med.uni-giessen.de
ORCID: 0000-0002-7876-4962
Computational Cognitive Neuroscience and Quantitative Psychiatry
Justus-Liebig-Universität Gießen, Germany

*Corresponding author:
Zhicheng Lin
zhichenglin@gmail.com
ORCID: 0000-0002-6864-6559
Department of Psychology, University of Science and Technology of China

**Author contributions**
Both authors contributed to the writing and editing of the manuscript. In early 2024, Z.L. mentioned writing a paper on AI-based coding in a podcast interview with T.Z. Coincidentally, T.Z. had drafted a manuscript on coding using LLMs. Subsequently, they combined the drafts each had written.

**Competing interests**
The authors declare no competing interests.

**AI usage**
We used GPT-4o and Claude 3.5 Sonnet for proofreading the manuscript, following the prompts described at https://www.nature.com/articles/s41551-024-01185-8.

**Acknowledgments**
Z.L. was supported by the National Key R&D Program of China STI2030 Major Projects (2021ZD0204200), the National Natural Science Foundation of China (32071045), and the Shenzhen Fundamental Research Program (JCYJ20210324134603010). T.Z. was funded by Deutsche Forschungsgemeinschaft (DFG, German Research Foundation) – SFB/TRR 135, sub-project C11.

**Abstract**
Computer programming (coding) is indispensable for researchers across disciplines, yet it remains challenging to learn and time-consuming to carry out. Generative AI, particularly large language models (LLMs), has the potential to transform coding into intuitive conversations, but best practices and effective workflows are only emerging. We dissect AI-based coding through three key lenses: the nature and role of LLMs in coding ("why"), six types of coding assistance they provide ("what"), and a five-step workflow in action with practical implementation strategies ("how"). Additionally, we address the limitations of AI in coding and explore its future outlook. By offering actionable insights, this framework helps guide researchers in effectively leveraging AI to enhance coding practices and education, accelerating scientific progress.

Coding skills are indispensable for researchers in biomedical sciences and, increasingly, social sciences. Yet learning to code is challenging—and coding, even for experienced programmers, can be time-consuming and dreadful. Many researchers code not because they enjoy it but because they need to get something done—such as conducting simulations; collecting, processing, and analyzing data; or visualizing results. Traditionally, researchers could delegate these tasks (e.g., by hiring a research assistant), use specialized packages (e.g., MATLAB Psychtoolbox for data collection, ggplot2 in R for visualization), adapt old code (e.g., from previous projects or someone else's, such as from Stack Overflow), or start from scratch. With the advent of generative artificial intelligence (AI), particularly large language models (LLMs), we now have a transformative new option: chatting with AI through natural language, which serves as a productivity and pedagogical tool with real benefits but also poses potential risks, such as hidden bugs or security threats[1-3].

Using LLMs for coding may seem straightforward—a conversation away. However, effectively leveraging these tools proves challenging, as AI-assisted coding represents a fundamentally new approach and requires proactive training and consideration[4-6]. It requires a different mindset, effective prompts[7,8,9], and a workflow tailored to LLMs. While best practices are beginning to emerge[10,11], such expertise—acquired through experience—is often implicit, sporadic, and not widely available. Urgently needed, then, is targeted training that informs productive integration and ethical use[3,4,6,12-16].

Based on our experiences as working scientists leveraging AI tools for coding, in this article we dissect the core strengths, utility, and potential of AI-assisted coding by 1) analyzing its nature and role in research and educational practices ("why"), 2) mapping types of AI assistance ("what"), and 3) outlining a five-step workflow ("how"; see also **Box 1**). To provide a concrete example, we detail how LLMs were used to make the code for **Figure 1** in the article. Additionally, we discuss how to report AI usage in coding; strategies for using and prompting LLMs for coding (**Box 2**); and AI-based coding in educational practices (**Box 3**). Limitations and future outlook are also evaluated. By providing actionable insights and strategies for leveraging AI in coding, we hope this article serves as a much-needed guide for researchers and instructors.

**Nature and role of LLMs in coding**
Programming languages were created to allow for precise, efficient, and sophisticated control over hardware. Over the years, they have gone through cycles of transformation, moving from low-level languages that require a deeper understanding of hardware (e.g., assembly languages), to more accessible, higher-level languages—languages with simpler syntax and more extensive built-in functions (e.g., Python). By abstracting away the complexities of machine code to focus more on concepts, design, high-level architecture, and problem-solving, high-level languages have made coding more accessible, thereby enabling more people to exploit these opportunities.

The recent emergence of LLMs in coding marks a transformative moment in the evolution of programming languages for three reasons. First, coding is now morphing into a more intuitive dialogue between the user and the AI. By reducing the need for syntax mastery, it brings computer instructions closer to everyday communication, allowing users to expend less effort on the mechanical and less enjoyable aspects of coding[3]—and more on strategic problem-solving and communicating and collaborating with AI. This reshapes the division of labor. Second, the vast knowledge embedded in LLMs allows AI systems to offer much more than just coding assistance—they help clarify concepts, brainstorm ideas, improve user understanding[17]. Third, the combination of their intelligent, versatile, and collaborative/interactive nature allows LLMs to offer personalized, on-demand assistance—for example, based on data from the user or tailored to the user's background, and with interactions that adapt to the user's skill levels and learning paces. AI tools thus fundamentally reshape our coding workflow, serving not just as tools but as collaborators—empowering novices and experienced programmers alike to tackle mundane tasks and new challenges with confidence and improved efficiency[16].

By making coding more accessible, efficient, and enjoyable, LLMs are democratizing coding skills on an unprecedented scale[18]. The implications are likely profound for educational and research practices[16,19,20]. This is because, despite the increasing importance of coding—driven in part by widespread data digitization and a push for open, reproducible science—systematic training is often lacking in academia[21]. Researchers, particularly those early in their careers, juggle learning multiple skills, and coding is often something to be learned on the job—mostly by oneself, with occasional assistance from labmates. The silent struggle can breed stress, anxiety, and self-doubt. LLM-based coding alleviates these challenges through personalized, on-demand assistance, opening new doors for learning and building confidence. Such tailored instruction is particularly effective for self-learning, offering a practical, scalable solution to Bloom's two-sigma problem (i.e., one-to-one tutoring leads to performance that is two standard deviations better than regular classroom settings)[22].

**Mapping types of AI assistance**
To illustrate how LLM-based coding addresses common research needs, we showcase use cases by categorizing them into six types of AI assistance: code understanding[17], code generation[23,24], code debugging[25], code optimization[26], code translation[27], and code learning[28].

*Code understanding*
A common challenge in research is comprehending and building on unfamiliar code—whether it is studying tutorials, examining code from another study, collaborating with other researchers, joining a new project, adapting legacy code, or even understanding our own code from a past project. As intelligent interpreters, LLMs excel at deciphering and explaining code, tailoring explanations to the user's experience and background.

For example, when studying complex tutorials, LLMs can link code implementations to underlying concepts and break down concepts step-by-step. For interpreting legacy code and code from published papers, submitted manuscripts, or collaborators, LLMs can provide high-level overviews of code structure, highlight key algorithms, identify key modules and their relationships, summarize the functionality of different components, provide detailed annotations, and explain the logic behind specific implementations. This capability extends to revisiting one's own code from past projects, where LLMs can refresh our memory and explain the rationale behind previous coding decisions.

*Code generation*
Writing code is bread and butter for many researchers—from stimulus presentation to data preprocessing, analysis, and visualization. LLMs can enhance productivity and code quality, allowing researchers to focus more on the scientific aspects rather than getting bogged down in coding technicalities.

Consider, for example, when starting a new project: LLMs can generate boilerplate code, suggest appropriate libraries/frameworks, or adapt preexisting code for the present project, thus helping overcome the "blank page" syndrome. They can also streamline time-consuming and repetitive tasks like data manipulation and preprocessing. For more complex tasks such as translating mathematical or conceptual algorithms into functional code, LLMs can assist by providing code snippets or full implementations of common algorithms—while the code may not work exactly as requested, it offers a convenient start for iteration. Perhaps the most ubiquitous use case is its utility in adding in-line comments and organizing/documenting code—essential for promoting research reproducibility and facilitating collaboration but time-consuming and not incentivized, and thus often overlooked in research practices.

*Code debugging*
Debugging can be tedious and frustrating. Error messages are often cryptic to inexperienced eyes. Common issues include identifying and fixing syntax errors and troubleshooting incorrect results or unexpected behavior. LLMs help with debugging by providing clear explanations of error messages, suggesting potential fixes, and offering strategies to resolve more complex issues. By highlighting common pitfalls and best practices, LLMs also contribute to the overall improvement of coding skills and practices.

For example, when encountering a syntax error, LLMs can break down the error message, explain the underlying syntax rule, and suggest corrections. For more complex issues like incorrect results or unexpected behavior, LLMs can analyze the code, identify potential problem areas, and suggest debugging strategies or alternative approaches. Fixing statistical models or simulations that yield anomalous results is particularly challenging, and LLMs can suggest modifications in the code or algorithm that align with the specific dataset and expected theoretical models.

*Code optimization*
Good code is not only functional but concise, readable, robust, and efficient. This is particularly vital when working with large datasets or complex algorithms, and when preparing code for publication or sharing with collaborators. Optimizations include adhering to best coding practices, enhancing code readability with descriptive variable/function names, maintaining consistent formatting and naming conventions, refactoring redundant code into functions, and improving algorithmic efficiency. LLMs enable researchers to refine their codebase effectively.

For instance, when optimizing code to better handle complex computations, LLMs can assist by analyzing the code and suggesting context-aware refinements that can reduce execution time and resource consumption. This might involve explaining the efficiency of various algorithms, suggesting improvements, or highlighting potential bottlenecks in the current code (e.g., replacing iterative loops with vectorized operations or modularizing frequently used code into functions).

*Code translation*

Translating code between programming languages is often needed to access specific libraries (e.g., a library available in R but not in Python), improve computational efficiency (e.g., from Python to C or Rust), minimize security vulnerabilities (e.g., from C to Rust), facilitate collaboration across teams that use different programming languages, or ease the transition to open-source alternatives (e.g., from MATLAB to Python or R). Code translation can be time-consuming and error-prone. LLMs excel at code translation, offering researchers a powerful tool to bridge these gaps.

For example, LLMs can assist by translating code from one language to another, suggesting equivalent libraries or functions, explaining necessary adaptations, and highlighting potential pitfalls or differences that may affect results. By more effectively tapping into our knowledge and experience, LLMs facilitate understanding and learning new programming languages.

*Code learning*
Learning and improving coding skills is an ongoing need as well as a challenge for many researchers. Common needs include understanding basic and new programming concepts, learning syntax and best practices, and applying coding skills to domain-specific problems. LLMs serve as a patient, always-available tutor that allows researchers to learn at their own pace, focus on concepts most relevant to their work, and gain hands-on experience with custom examples. Additionally, drawing from vast resources on the internet, LLMs can design customized courses for individual users. This targeted approach not only enhances learning efficiency but also boosts confidence and joy. LLMs thus allow users to conveniently access highly relevant knowledge, making learning more efficient and aligned with their goals.

For instance, to facilitate learning, LLMs can provide varied domain-specific worked examples. By using neural data for a neuroscientist and financial datasets for an economist, this context-aware approach helps bridge the gap between abstract coding concepts and practical, relevant applications. To reinforce learning and test understanding, LLMs also excel at generating tailored exercises based on the researcher's progress and areas of difficulty—for example, creating a series of increasingly complex exercises and providing immediate feedback for each attempt. Furthermore, as researchers work through coding challenges, LLMs can offer real-time problem-solving strategies: breaking down complex tasks into manageable steps, suggesting alternative approaches, and explaining the pros and cons of different solutions.

**A five-step workflow for AI-assisted coding**
To facilitate effective integration of LLMs into coding routines, below we outline a general five-step workflow that serves as a working framework (**Figure 1**). The workflow can be adapted to different needs—for example, a one-off, simple request would require minimal preparation, but the general procedure, guiding principles, and practical wisdom can be equally useful. A summary version is available in **Box 1** for easy reference.

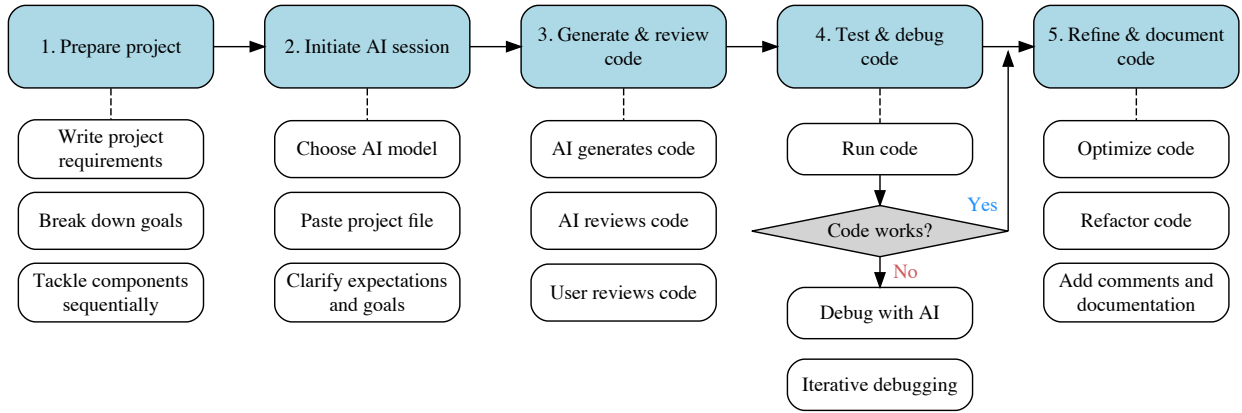

**Figure 1 | Five-step workflow for AI-based coding in scientific research.** The five-step process for integrating LLMs into coding tasks starts with project preparation, followed by initiating an AI session, generating and reviewing code, testing and debugging the code, and finally refining and documenting the code. The workflow emphasizes the importance of specifying the project's requirements and goals, as well as the iterative nature of coding. This structured approach provides a general and practical framework for researchers to effectively leverage AI in their coding practices.

### *Step 1: Prepare your project*
Effective AI-assisted coding begins with project preparation. To ensure that AI outputs align with your goals, start by describing your project requirements and purpose in a document. Include the relevant context of your project, the problems you aim to solve, and your constraints or preferences (e.g., preferred language or libraries). This document serves as a roadmap for the project, helping you clarify your thoughts and keeping the AI focused on your needs. Save it for reference and for sharing with collaborators or the AI.

To enable more precise, targeted interactions with the AI, consider breaking down higher-level goals into smaller, well-defined steps. For example, instead of a broad directive like "analyze my dataset", break it down into steps: "load data", "clean missing values", "perform descriptive statistics", and "visualize the results".

For larger, more complex projects, tackle each component separately. For instance, to develop a comprehensive data analysis workflow, consider decomposing it into separate components: data acquisition, preprocessing, analysis, and visualization. Complete each component, and then start a new session to continue with the next. This approach helps maintain clarity and prevents overwhelming the AI with too much context.

### *Step 2: Initiate the AI session*
The next step is to select an appropriate AI model. Major AI coding tools are organized into four categories in **Table 1**, including generalist LLMs, plugins, code editors, and notebook platforms. Which tools to use depends on personal preferences and needs. Consider using advanced models such as Claude 3.5. These models offer robust capabilities for coding in various programming languages.

Next, share your prepared project document with the AI by copying and pasting the text directly into the chat interface or uploading the document. Before proceeding with specific coding tasks, ask whether it needs any additional information or clarification about your project. Confirm its understanding, and encourage it to ask questions if anything is unclear.

**Table 1 | Major AI coding tools as of October 2024**

| Tool | Nature | Utility | Link |
| --- | --- | --- | --- |
| Claude 3.5 Sonnet, GPT 4o, Gemini 1.5 Pro, Llama 3.2, DeepSeek-V2.5 | Generalist LLMs (some also with specialized features)* | Provide coding assistance and general-purpose and contextual support | https://www.anthropic.com <br> https://openai.com <br> https://gemini.google.com <br> https://aistudio.google.com <br> https://ai.meta.com <br> https://www.deepseek.com |
| GitHub Copilot, Cody | Plugins for code editors (e.g., VS Code) and IDEs (e.g., Visual Studio) | Provide code suggestions, autocomplete, and, for Cody, other assistance like refactoring and explanations to help improve coding speed and fluency | https://github.com/features/copilot <br> https://sourcegraph.com/cody |
| Cursor, Zed, Replit | Code editors that integrate LLMs | Provide advanced coding support (e.g., generation, completion, refactoring, and debugging) directly within the editing environment | https://www.cursor.com <br> https://zed.dev <br> https://replit.com |
| Google Colab, Jupyter Notebook | Notebook platforms that either integrate AI features (Colab) or support AI extensions (Jupyter AI) | Provide interactive and collaborative coding environment with AI features (e.g., generation and completion) | https://colab.research.google.com <br> https://jupyter.org |

Note. VS Code is short for Visual Studio Code; IDEs for integrated development environments; LLMs for large language models.

* For a list of leading LLMs for coding, see: https://chat.lmsys.org/?leaderboard (select "Coding" from the "Category" dropdown menu) or https://bigcode-bench.github.io. Specialized

features include ChatGPT's Data Analyst (for data-related tasks; https://chatgpt.com/g/g-HMNcP6w7d-data-analyst) and Claude's Artifacts (for interactive visualizations and applications; https://support.anthropic.com/en/articles/9487310), which are good for quick prototyping and explorations (a key utility of Google Colab/Jupyter Notebook; see also **Box 2**).

*Step 3: Generate and review the initial code*
You are now ready to generate code. Be specific in your prompt, referencing the project document from Step 1 when appropriate (e.g., "*Create R code to perform descriptive statistics on the dataset as detailed in my project outline*"). For strategies for prompting and using LLMs for coding, see **Box 2**. Once you have the initial code, you can ask the AI to review it and suggest improvements—LLMs do not read or reflect on their outputs unless directly instructed to do so.

For a more robust review process, consider using multiple AI models. For instance, you could use Claude 3.5 Sonnet to detect errors or suggest optimizations, and then have ChatGPT-4 assess these suggestions. Return to Claude 3.5 Sonnet with the reviewed feedback for further assessment. This multi-model approach mimics the diversity of human feedback on platforms like Stack Overflow, which is instrumental for more comprehensive, reliable code improvements.

With the completed code, consider asking the AI for a step-by-step explanation of how the code accomplishes your stated goals (e.g., "*Walk me through how this code performs the descriptive statistics analysis*"). To further verify its alignment with your request, start a new AI session, paste the generated code (without any context), and ask for an explanation of what the code does. If the explanation accurately recreates your original intent, it is a good indication of code accuracy and alignment.

Whether you have the AI review the code or not, in general you should aim to understand the code by examining it line by line. A good grasp of coding is beneficial, but AI can also help this process by providing comments and explanations for each line of code. For larger, more complex code, divide it into logical segments, and try to understand each segment before moving on to the next. For one-off tasks where the goal is simply to get something done quickly (e.g., an intermediate step of data format transformation) and the code is self-contained (that is, not part of a codebase), code quality is not critical and a line-to-line level of understanding not practical.

*Step 4: Test and debug the code*
The next step is to ensure that the code functions as intended. Run the code. If the code fails to run, share the results and error messages with the AI for debugging suggestions. If the code runs but produces unexpected outcomes, provide the AI with a detailed description of the expected versus actual results, and check the assumptions made in the code or the algorithms used. Give it clear feedback. Critically evaluate the suggestions from the AI using your domain knowledge and your understanding of the project requirements. Trust your intuition. To maximize efficiency, focus on getting the core features of your code working first—leave minor bug fixes and additional features for later.

Debugging is often a process. If the initial fixes suggested by the AI could not resolve the issue, paste the outcomes to the AI again and request alternative solutions. If stuck, consider adjusting the AI settings (such as the temperature parameter in GPT-4) or starting a fresh conversation. By removing the context of the original conversation, a new conversation promotes a different perspective. Consider also alternating between different AI models, such as ChatGPT-4 and Claude 3.5 Sonnet, as each model shares different strengths and limitations—leveraging multiple models can help identify and resolve a wider range of issues. Take breaks between debugging sessions—the performance of LLMs can vary even with the same prompt.

For code that involves data analysis, it also helps to validate the code with known data and results. A small, representative dataset suffices. By comparing the output from the code with the known correct output, we can verify the code's accuracy with more confidence.

However, it is important to set realistic expectations: LLMs will not solve all coding bugs. Understanding their limitations—in debugging and beyond—is part of learning to use LLMs effectively. If several rounds of back-and-forth fail to resolve a bug, it may be time to try other approaches. With experience, you will develop an intuition about when to consult LLMs and when to Google instead.

### *Step 5: Refine and document the code*
With working code in hand, now is the time to refine and document it—making it efficient, maintainable, and understandable to others (including your future self!).

Begin by optimizing for performance and efficiency (*"Review the code and suggest ways to improve its performance, particularly focusing on [specific areas of concern]"*). Ask the AI to refactor the code to enhance its structure and readability (*"Refactor the code to improve modularity and simplify its design without altering its functionality"*). Implement the suggestions incrementally and test the code (following Step 4) to ensure each modification maintains functionality while improving performance/readability.

Next, ask the AI to add in-line comments explaining how each section of the code works (*"Add detailed in-line comments to the code, explaining the purpose and functionality of each section"*). Additionally, ask the AI to generate overall documentation for your project (*"Create a project overview document that includes a description of the project's purpose, how to run the code, and any dependencies or setup requirements."*)

**AI-based coding in action**
To showcase AI-based coding in action, we describe how author Z.L. used LLMs to generate the code for **Figure 1**, following the five-step framework, with reflections on key design choices. Due to the personal nature of this process, the pronoun "I" will be used.

### *Step 1: Prepare your project*
As part of the preparations, I had the manuscript ready, with a clear goal: to draw a flowchart of the five-step workflow. I also sketched a simple flowchart design on a tablet (Kindle Scribe), using five rectangles to represent the steps, arranged horizontally and connected by arrows.

While I knew how to use the DiagrammeR library in R to draw flowcharts, it had been quite a while since I last used it. I decided to use LLMs to jumpstart the project. In the end, drawing the figure took about three hours, and I estimated that without AI assistance, it would probably have taken twice as long. This is an important point to emphasize: AI-based coding is not just for beginners, but it can also save time for more experienced coders.

*Step 2: Initiate the AI session*
I decided to use Claude 3.5 Sonnet first because it is one of the best LLMs for coding. Additionally, it had the Artifact feature, which displayed the results of code execution instantly. This immediate visualization is helpful as it facilitates quick prototyping.

*Step 3: Generate and review the initial code*
As shown in **Figure 2**, I started by uploading the manuscript and entering the following prompt: "*Show me a flowchart for the AI human coding workflow from the manuscript, using Python*". Claude responded by explaining the tool used ("Mermaid") and detailing the content of the flowchart ("five-step workflow outlined in Box 1"), along with providing the code and a preview of the flowchart.

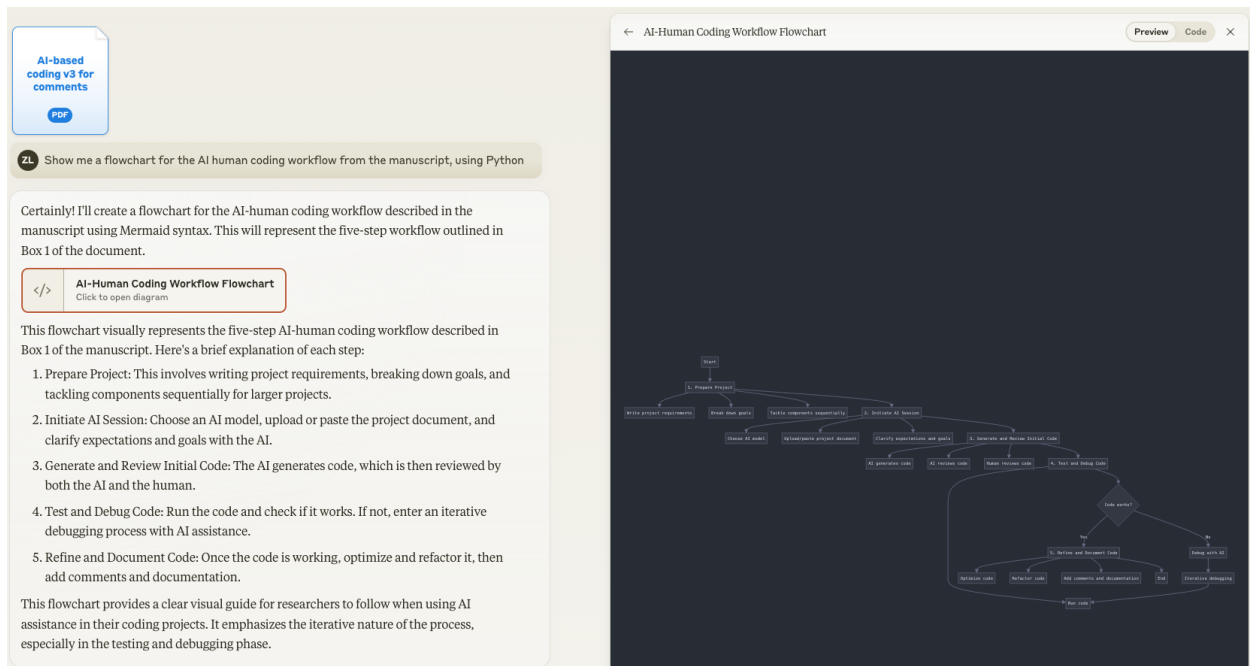

**Figure 2 | First interaction with Claude.** The Python code for the flowchart was successfully generated and visualized (right) based on the uploaded manuscript and a simple prompt (left).

The flowchart required modification. To steer it in the right direction, my first idea was to include the substeps within the box of the main step. So I followed up with the following prompt:

"*OK. Looks promising, but needs some work. Put the components within each step rather than outside of the component. For example, 1. Prepare Project, you can put the three boxes of Write..., Break..., and Tackle... into that.*"

However, the revised flowchart was not quite what I looked for (**Figure 3**).

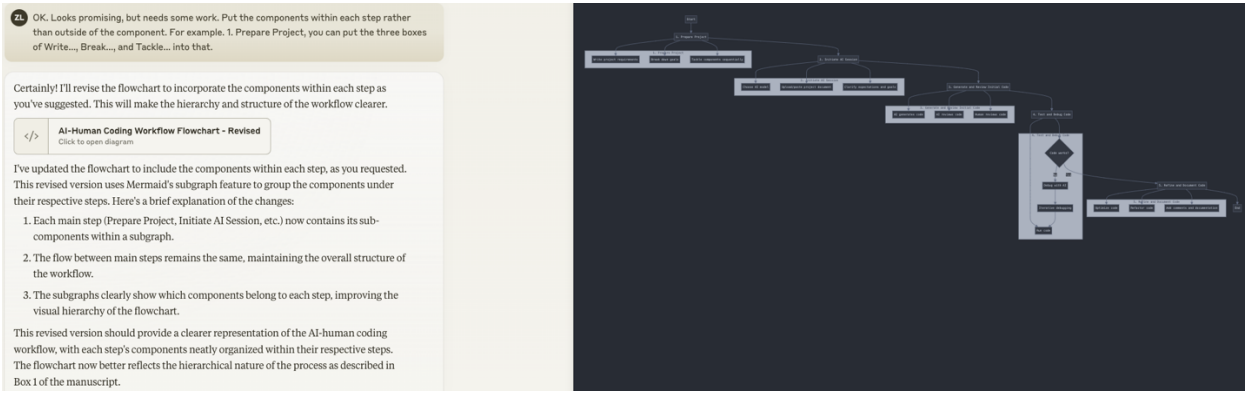

**Figure 3 | Second interaction with Claude.** The first attempt to rearrange the items did not yield satisfactory results.

So I decided to edit the prompt to be more specific:

"*OK. Looks promising, but needs some work.
1: Remove the "Start"
2: Arrange the five components horizontally, from left to right, 1 to 5
3: For each component, arrange the substeps under the component and vertically (up to down). For example, 1. Prepare Project, you can put the three boxes of Write..., Break..., and Tackle... below it, and arrange them vertically.*"

But the spatial arrangement of the flowchart was still off (**Figure 4**).

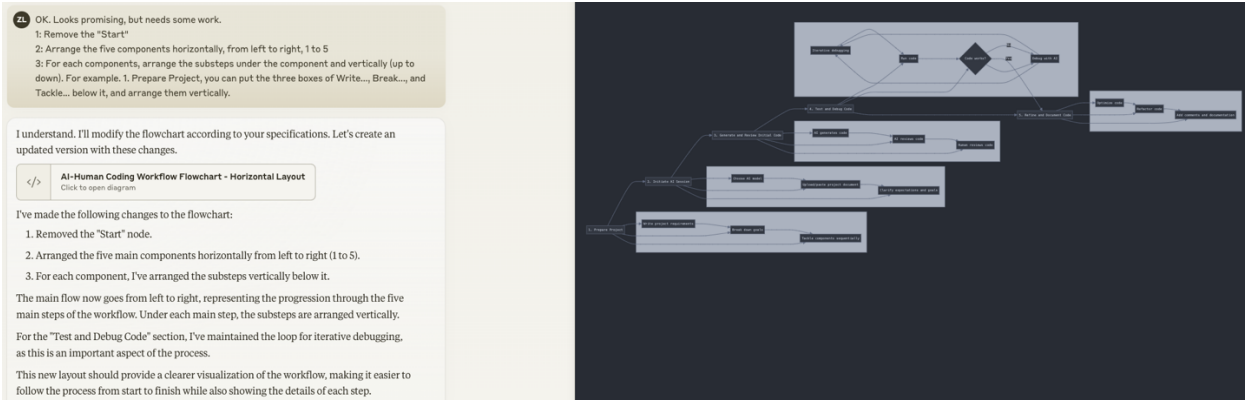

**Figure 4 | Third interaction with Claude.** The second attempt to rearrange the items, by making the prompt more specific, also did not yield satisfactory results.

I decided to edit the attempt further, by using spatial coordinates:

"*OK. Looks promising, but needs some work.
1: Remove the "Start"
2: Arrange the five components horizontally, from left to right, 1 to 5. Remember, they need to have the same y value, meaning they are arranged within a straight horizontal line*

*3: For each component, arrange the substeps under the component and vertically (up to down), with the same x value. For example. 1. Prepare Project, you can put the three boxes of Write..., Break..., and Tackle... below it, and arrange them vertically."*

While the updated flowchart looked somewhat better, the spatial arrangements were still not right (**Figure 5**). So, I decided to stop here, and iterate from the final Python code provided by Claude.

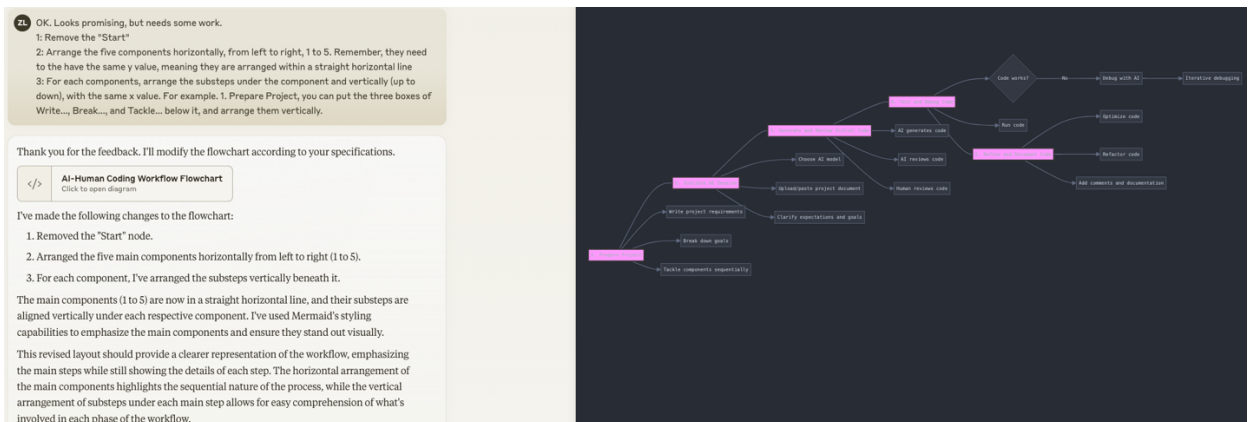

**Figure 5 | Fourth interaction with Claude.** The third attempt to rearrange the items, by specifying spatial coordinates, improved the design but still did not yield satisfactory results.

*Step 4: Test and debug the code*
I was not familiar with Python, so I decided to translate it into R code first using ChatGPT 4o, and iterate from there. The entire chat history is available here: https://chatgpt.com/share/279a99dd-9f3c-48a7-8578-a8c4f6ca99ca.

Two points about the different strengths of LLMs:
- Unlike ChatGPT, Claude did not support chat history sharing at the time of writing;
- Unlike Claude 3.5 Sonnet, ChatGPT 4o did not natively support code execution (available only through a special GPT called Data Analyst). But this was fine by me as I had RStudio installed, ready for testing and debugging.

Since the entire chat history is available, I will not detail my interactions but will provide a general summary. I asked ChatGPT to:
- Translate the Python code into R code ("*Convert the code to R*"). It managed this, but the exact item arrangements were a bit off (**Figure 6**).
- Make spatial rearrangements, based on the previous prompt for Claude (see **Figure 5**), which yielded much better results than Claude (**Figure 7**).
- Make specific adjustments, including making the five rectangles the same size ("*For the five main steps, can you make the rectangle the same size? Only show me the updated code.*"), improving text alignment ("*Update the code below so that the text is the same length (if too long, break them into two lines):*"), rounding the boxes, and changing alignments, size, and other visual features to my liking.

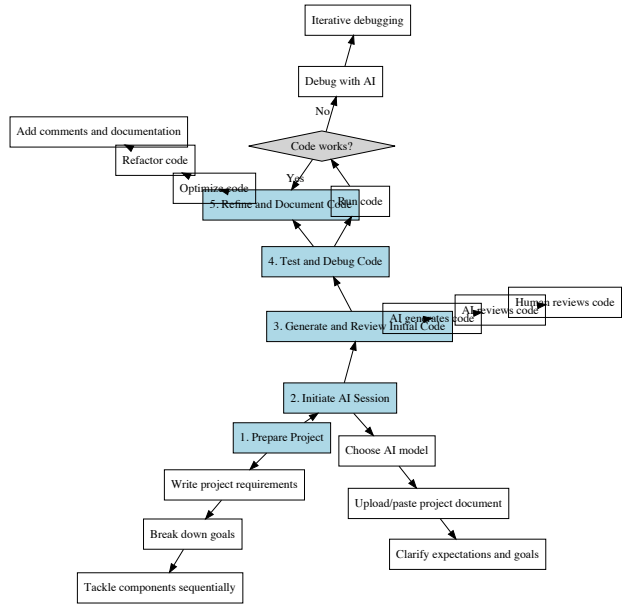

**Figure 6 | Results of R code after translation from Python using ChatGPT-4o.** The translated code successfully produced a flowchart as well, though the spatial arrangements differed from the original flowchart in Python.

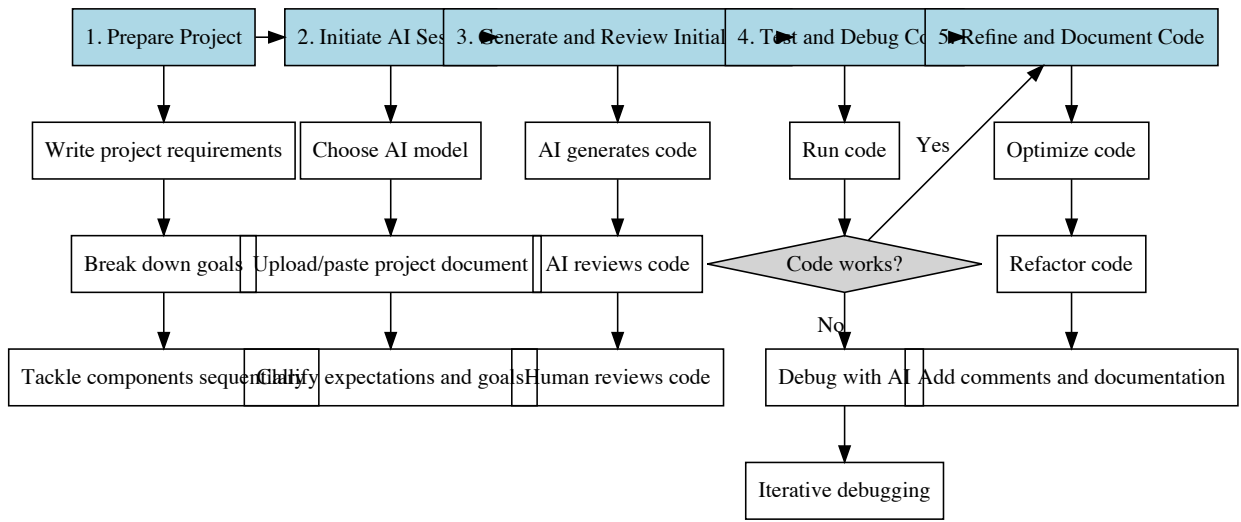

**Figure 7 | Spatial rearrangements using ChatGPT-4o.** Using the same prompt to rearrange the items now showed much better results; the structure was very close to the requested arrangements, and only some minor adjustments were needed.

*Step 5: Refine and document the code*
I was able to adjust the spatial arrangements and visual features to my liking using ChatGPT-4o (**Figure 1**). I just needed to export the figure to a PDF file, so I borrowed code from my previous projects. The final code produced the figure I wanted, and the code looked well-documented and clear

(https://www.dropbox.com/scl/fi/an4ukijwscwxb0ydchbgm/plotCoding.R?rlkey=bqwfafs0a18lsp2cqsiui04db&dl=0).

**Reporting the usage of AI tools in coding**
To maintain transparency, reproducibility, and compliance with institutional policies, we need to document AI assistance in coding. However, this is more challenging than it sounds: journal policies are often unclear or inconsistent; full documentation is cumbersome if not impractical; and complete reproducibility at the code/text generation level is not feasible[29]. The bottom line is to follow relevant policies—journal policies for publications and university policies for teaching and research. But this needs to be a forethought rather than an afterthought: while most policies generally do not forbid the use of generative AI, variations do exist across different policies and contexts.

Best practices are still emerging. To balance transparency and practicality, a starting point is to prioritize the documentation of AI tools (including name, version, and time), their roles in coding projects, and specific prompts and interactions. Most journal policies require AI usage to be disclosed in the methods or acknowledgment sections. Yet these policies mostly address text writing rather than coding. Thus, for code documentation, we additionally suggest including a section in the readme file or at the beginning of the code file to document the usage of AI tools. And if needed, use in-line code comments to highlight sections of code that were generated or altered by AI tools.

**Limitations**
Despite the transformative potential of LLMs for coding in research, there are important limitations[3,12]. As with other applications, LLMs can hallucinate, such as suggesting non-existing functions or libraries[9]. In coding contexts, this issue is mitigated to some extent because the output can be verified by running the code. Another limitation is that, due to uneven training data, LLMs perform better in some languages (e.g., Python) and tasks (e.g., modular, classic/common ones) than others (e.g., NetLogo; unfamiliar problems)[30]. Expanding training datasets will be crucial to address these limitations. More generally, while LLMs excel at language and pattern recognition, they currently fall short of human qualities like intuition, common sense, and contextual understanding and reasoning. This limitation means that, for example, if LLMs cannot fix a bug after two to three tries, it is unlikely that they will succeed with additional attempts, showing a limit in iterative troubleshooting[30]. Ethical issues also arise with AI-assisted coding, including equitable access to AI coding tools and transparency about their use in publications[29].

While LLMs lower the barrier to entry for coding and improve the efficiency of learning and coding, they do not eliminate the need for foundational knowledge and skills[6]. In a 1995 interview, Steve Jobs commented that "everybody should learn to program a computer, because it teaches you how to think". A basic understanding of programming concepts, familiarity with relevant packages and libraries, domain-specific knowledge, critical thinking—all of these help researchers ask better questions, select appropriate tools, and evaluate AI-generated code, particularly in distinguishing between correct versus plausible-looking but incorrect outputs.

Indeed, one danger of relying on AI-generated code is that insecure or incorrect code becomes integrated into the codebase simply because it produces outputs that appear correct[6]. Without understanding the code, this codebase may become too unwieldy to debug or modify. Additionally, over-reliance on LLMs may lead to a decline in certain coding skills (e.g., writing from a blank page), coding intuition, and peer collaboration for lack of practice[6]. And when we come to rely on AI to correct or optimize our code, it may also lead to decreased attentiveness to code quality[14]. Be vigilant, then, of AI mistakes and biases as well as the danger of over-relying on AI.

**Outlook and concluding remarks**
Computer languages represent a most fertile ground for automation: they are generally more logical than natural languages; there is a large corpus of high-quality code available for training; and there is a strong demand, both commercial and otherwise. AI agents are being developed to automate aspects of coding tasks—a development likely to further transform the landscape of software development and scientific research. The future of abundant, low-cost, high-quality software is thus poised to accelerate a wide range of creative endeavors—as well as raise new ethical and philosophical questions (for example, if our productivity greatly depends on AI, are we still coders, or are we more like proficient tool users, and does this matter to our identity?).

As illustrated in this article, LLMs and various AI tools right now are already very useful for coding. By bridging natural language and computer instructions, LLMs are reshaping how researchers approach computational tasks and redefining the role of coders and our workflow. In some sense, proficient AI-based coding requires the mindset of a good designer: focusing not just on the appearance of the code but on its functionality—how it operates and solves problems. By democratizing coding skills and offering personalized, on-demand assistance, LLMs have the potential to fundamentally transform scientific coding practices—serving as both tools and collaborators. Yet, building a symbiotic relationship with AI requires that we develop a new set of AI skills to become better designers: the ability to see the big picture and communicate with AI, evaluate its outputs, and integrate AI into our workflows. It also requires curricula to adapt and prepare students for this AI-infused future (see **Box 3** for an outline to integrate AI-based coding into the classroom). The benefits are not just cognitive, enhancing efficiency and productivity; they are also emotional, reducing distractions and bringing confidence and joy.

**Box 1 | A five-step workflow for AI-assisted coding**

1. **Prepare your project**
   - **Guidance:** Write out your project requirements and goals in a document; save this document for reference and sharing.
   - **Steps:** Break higher-level goals into smaller, well-defined steps.
   - **Components:** Tackle larger projects by focusing on individual components sequentially.

2. **Initiate the AI session**
   - **Choose AI model:** Consider Claude 3.5 Sonnet, ChatGPT-4o, or Llama 3.2.
   - **Initiate:** Upload or paste your document into the AI interface.
   - **Clarify expectations and goals**: Ask the AI if it needs any additional information and confirm the AI understands what you want it to do.

3. **Generate and review the initial code**
   - **Generate:** Have the AI generate the initial code based on your requirements.
   - **AI review:** Ask the AI to review the code and make corrections that make sense (consider leveraging multiple AI models); ask the AI to walk through step-by-step how each piece of code accomplishes your stated goals; start a new session and paste the code into AI, asking it to explain again what the code does step-by-step, and ensure that the explanation matches the original requirements.
   - **Human review:** Try to understand the structure and function of each line of code.

4. **Test and debug the code**
   - **Test:** Run the code to see actual results; for data analysis, consider validating with known data and results
   - **Debug:** If the code does not work, provide the AI with the execution results to suggest fixes; if the code runs but the results do not look right, examine the assumptions made in the code or algorithms used.
   - **Iterative debugging:** If the initial fixes are not successful, request alternative solutions; take breaks as needed; if stuck, adjust the settings (e.g., temperature in GPT-4) or start a new conversation; focus on getting the core features working first; consider alternating between different models to identify and fix bugs.

5. **Refine and document the code**
   - **Optimize:** Request the AI to refine and optimize the code for performance improvements and efficiency (test the code by following Step 4 above).
   - **Refactor:** Ask the AI to refactor the code to simplify the design (test the code by following Step 4 above).
   - **Document:** Request the AI to add in-line comments explaining how the code works and provide documentation (e.g., project overview, how to run the code).

**Box 2 | Strategies for using and prompting LLMs for coding**
LLMs do not always provide satisfactory answers on the first attempt—or even after several attempts. Iterations may be needed to fine-tune the results. But to maximize efficiency and minimize iterations, it pays to learn to create more effective prompts. Below, we outline four strategies in the context of coding.

**Strategy 1: Play with various LLMs**
No strategy is more important than experimenting and playing with various LLMs and their specialized features. Such experience helps build intuitions about what the LLM is capable and not capable of, what kinds of mistakes it makes, and how specific our instructions need to be. For some seemingly complex tasks, a simple prompt may suffice (e.g., tasks that are well documented on the internet); for others, we need to break them down into clear steps. Sometimes our prompt needs to be explicit (e.g., spelling out an acronym that can be ambiguous in the conversation), but sometimes a succinct one is all the model needs (e.g., "*Use ggplot2 to ...*"). Some bugs may arise from model hallucinations (e.g., a nonexistent function) and yet some may be due to typos in our input (e.g., variable names). These intuitions evolve as models are continuously being updated—but they can only be gained through hands-on experiences.

While experimenting with LLMs, play with their specialized features too. For instance, a custom version of ChatGPT is Data Analyst, which we can directly upload a dataset to and chat with the data—compute mean and standard deviation, visualize these metrics, and more. Similarly, Claude's Artifacts is a feature that enables interactive visualizations and applications. Both can facilitate rapid prototyping akin to Jupyter Notebook. By playing with various LLMs and their specialized features, we can develop a better sense of their capabilities and limitations.

**Strategy 2: Provide clear and detailed instructions**
To steer the model toward our desired outcome, it helps to be specific and informative in our instruction, such as the desired code language, the input formats, desired outputs, libraries, frameworks, or any other relevant details. Reiterating the instruction at the end of the prompt can also make it more likely that the LLMs adhere to the instruction throughout the task. With ambiguous instructions, LLMs resort to making assumptions. Thus, explicit, clear, and detailed instructions promote accurate and relevant responses from LLMs.

Sometimes this means steering the model toward specific, concrete coding tasks, such as individual functions as opposed to an entire analysis pipeline. If unsure about how to outline and break down tasks, we can also ask the LLM to provide a step-by-step outline and iterate from there.

Thus compare two prompts:
- Vague request: "*Describe the data in python*"
- Detailed instruction: "*Write a Python script that reads a CSV file of student grades, calculates the mean, median, and standard deviation for each subject, and outputs these statistics in a table. Then, generate a bar chart comparing the average scores in math and physics, labeling the mean on top of each bar. Use pandas for data manipulation and matplotlib for visualization.*"

**Strategy 3: Provide relevant context and goals**
Providing background and clear objectives helps the model understand the purpose and scope of our request, focus on the desired outcomes, and minimize the risk of irrelevant or off-target responses.

Thus compare:
- Vague request: "*t-test in Python.*"
- Instruction with relevant context and goals: "*The goal is to compare the results of two different conditions within subjects and determine whether the differences between conditions are statistically significant. In a CSV file, the data is stored in two lists, each representing the scores under Condition 1 and Condition 2 for the same group of subjects. Load the data, perform the paired t-test, and output the test statistic and p-value. Please offer a Python code using scipy.stats.*"

**Strategy 4: Give examples**
Examples allow LLMs to emulate the context, style, and structure of our code, thereby aligning them with our approach to problem-solving. Such examples can come from tutorial pipelines, scripts from previous projects, or one we work out ourselves, ensuring that the LLM produces results that are more accurate and relevant. An example prompt that embeds this approach is:

"*Below is some tutorial code from Python-MNE (https://www.frontiersin.org/journals/neuroscience/articles/10.3389/fnins.2013.00267/full) demonstrating independent components analysis (ICA) for magnetoencephalography (MEG) preprocessing:*

*import numpy as np*
*import mne*

*# Load data*
*sample_data_folder = mne.datasets.sample.data_path()*
*sample_data_raw_file = (*
   *sample_data_folder / "MEG" / "sample" / "sample_audvis_filt-0-40_raw.fif"*
*)*
*raw = mne.io.read_raw_fif(sample_data_raw_file)*

*# Set up and fit ICA*
*ica = mne.preprocessing.ICA(n_components=20, random_state=97, max_iter=800)*
*ica.fit(raw)*
*ica.exclude = [1, 2]  # Details on how we picked these are omitted here*
*ica.plot_properties(raw, picks=ica.exclude)*

*I am working on an Electroencephalography (EEG) preprocessing project with data named 'visual_info_raw.fif'. Please adapt the above code to suit my project.*"

**Box 3 | Integrating AI-based coding into the classroom**

AI-based coding offers new opportunities for enhancing student engagement and learning outcomes. This box provides an outline to incorporating AI tools into coding education.

**1. Coding and the role of AI**

**1.1 Nature of coding**
- Introduce students to the fundamental principles of coding, including logic, syntax, algorithms, and problem-solving.
- Explain the basic building blocks of coding, such as variables, control structures, functions, and data structures.

**1.2 Traditional coding versus AI-based coding**
- Compare traditional coding methods and AI-based coding, highlighting the strengths and limitations of each approach.
- Emphasize the conceptual shift from writing code manually to interacting with AI models to generate, debug, and optimize code, emphasizing how AI tools can serve as both collaborators and learning aids, particularly for novice coders.
- Discuss the specific advantages of using AI for coding, such as accessibility, reduced coding time, enhanced problem-solving capabilities, and the ability to quickly iterate and test code.

**Key point**: Emphasize the importance of a programming mindset and discuss the foundational knowledge students need to effectively use AI for coding. This foundation ensures that students are not merely relying on AI outputs but are also critically engaging with the content.

**2. The workflow of AI-based coding**

**2.1 The five-step workflow for AI-based coding**
- Present the five-step workflow outlined in the paper, which includes project preparation, AI interaction, code generation, debugging, and refinement.
- Provide detailed examples and refer to **Boxes 1** and **2** for additional strategies and tools that can enhance the learning experience.

**2.2 Hands-on practice**
- Assign specific coding tasks that require students to use AI assistance. These tasks should vary in complexity to cater to different skill levels, encouraging students to experiment with various prompts and explore the capabilities of different AI models.
- Encourage students to experiment with different AI models and prompting techniques. This exploration helps them understand the nuances of AI interactions and develop a deeper understanding of how to effectively leverage AI tools for coding.
- Introduce AI policies and how to properly report AI usage in coding.

**2.3 Peer learning and presentations**
- Select students who excel in their tasks to present their approaches to the class. This peer-learning strategy fosters a collaborative learning environment and allows students to learn from each other's experiences with AI-based coding.

- Promote the sharing of best practices and strategies for using AI tools. Encourage discussions around the challenges faced during AI-based coding and how these challenges were overcome, further enhancing the collective learning experience.

**Key point**: Introduce practical methods for AI-based coding and encourage students to engage in practice. By exploring coding through AI independently, students develop confidence and proficiency in using these tools.

### 3. Limitations, ethics, and future outlook
### 3.1 Limitations and ethical considerations
- Discuss the current limitations of AI in coding, including the potential for over-reliance and the need for foundational coding knowledge. Address ethical considerations, such as the importance of understanding AI outputs and the risks associated with blindly accepting AI-generated solutions.

### 3.2 The future role of AI in coding education
- Explore the potential future applications of AI in coding education and other domains. Encourage students to think critically about how AI can be integrated into their future work and studies, and how they can effectively collaborate with AI tools to enhance their productivity and learning.

**Key point**: By better understanding AI-based coding, students gain a broader perspective on how AI can enhance learning and productivity. This understanding can empower them to effectively collaborate with AI in their studies and future work.